\documentclass[%
 reprint,
 amsmath,
 amssymb,
 aps,
 pra,
twocolumn,
]{revtex4-1}

\usepackage{graphicx}
\usepackage{dcolumn}
\usepackage{bm}
\usepackage{gensymb}
\usepackage{physics}
\usepackage{color}
\usepackage{tikz}
\usepackage{soul}
\usepackage{upgreek}
\usepackage{makecell}
\usepackage{multirow}
\usepackage{hyperref}
\usepackage{subcaption}
\usepackage{scrextend}

\hypersetup{
    colorlinks=true,       
    linkcolor=blue,          
    citecolor=blue,        
    filecolor=blue,      
    urlcolor=blue           
}

\graphicspath{{./SrPAS-Figures/}}

\usepackage{txfonts}

\newcommand{\sr}[1]{\phantom{ }^{#1}\text{Sr}}

\newcommand{\term}[3]{\phantom{ }^{#1}\mathrm{#2}_{#3}}

\def\um{\mbox{ $\mu$m}}
\def\nm{\mbox{ nm}}

\def\kHz{\mbox{ kHz}}

\newcolumntype{.}{D{.}{.}{-1}}
\newcolumntype{d}[1]{D{.}{.}{#1}}

\newcommand{\lopt}{\ell_{\mathrm{opt}}}

\newcommand\T{\rule{0pt}{2.6ex}}       

\newcommand{\etal}{\textit{et al.\@}}

\begin{document}

\preprint{APS/123-QED}

\title{Narrow-line photoassociation spectroscopy and mass-scaling of bosonic strontium}

\author{B.J. Reschovsky}
\email{breschov@umd.edu}
\author{B.P. Ruzic}
\author{H. Miyake}
\author{N.C. Pisenti}
\author{P.S. Julienne}
\author{G.K. Campbell}

\affiliation{%
 Joint Quantum Institute, University of Maryland and National Institute of Standards and Technology, College Park, MD 20742
}%

\date{\today}

\begin{abstract}

Using new experimental measurements of photoassociation resonances near the $^1\mathrm{S}_0 \rightarrow \phantom{ }^3\mathrm{P}_1$ intercombination transition in $^{84}$Sr and $^{86}$Sr, we present an updated study into the mass-scaling behavior of bosonic strontium dimers. 
A previous mass-scaling model [Borkowski \textit{et al.}, Phys.\ Rev.\ A \textbf{90}, 032713 (2014)] was able to incorporate a large number of photoassociation resonances for $^{88}$Sr, but at the time only a handful of resonances close to the dissociation limit were known for $^{84}$Sr and $^{86}$Sr.
In this work, we perform a more thorough measurement of $^{84}$Sr and $^{86}$Sr bound states, identifying multiple new resonances at deeper binding energies out to $E/h=-5$~GHz. 
We also identify several previously measured resonances that cannot be experimentally reproduced and provide alternative binding energies instead.
With this improved spectrum, we develop a mass-scaled model that reproduces the observed binding energies of $^{86}$Sr and $^{88}$Sr to within 1~MHz. 
In order to accurately reproduce the deeper bound states, our model includes a second $1_u$ channel and more faithfully reproduces the depth of the potential.
As determined by the previous mass-scaling study, $^{84}$Sr $0_u^+$ levels are strongly perturbed by the avoided crossing between the $^1\mathrm{S}_0 + \phantom{ }^3\mathrm{P}_1$ $0_u^+$ $(^3\Pi_u)$ and $^1\mathrm{S}_0 + \phantom{ }^1\mathrm{D}_2$ $0_u^+$ $(^1\Sigma_u^+)$ potential curves and therefore are not included in this mass-scaled model, but are accurately reproduced using an isotope-specific model with slightly different quantum defect parameters. 
In addition, the optical lengths of the $^{84}$Sr $0_u^+,\ \nu=-2$ to $\nu=-5$ states are measured and compared to numerical estimates to characterize their use as optical Feshbach resonances.


\end{abstract}

\maketitle


\section{\label{sec:intro}Introduction}

Photoassociation is a process that couples two colliding ground-state atoms to a weakly-bound molecular state corresponding to one excited and one ground state atom \cite{Jones2006}. 
The process of measuring these bound states, known as photoassociation spectroscopy (PAS), can be used to probe the shape of the excited state molecular potential. 
Recently, PAS studies relative to the narrow $\term{1}{S}{0} \rightarrow \term{3}{P}{1}$ transition in alkaline-earth (like) atoms such as strontium, ytterbium, and calcium have been performed to measure these lines with precisions of $\lesssim 10$~kHz \cite{Stellmer2012,Borkowski2017,Kahmann2014a}.
Previous narrow line photoassociation spectroscopy (PAS) has been performed in $\sr{88}$ \cite{Zelevinsky2006,Mcguyer2013}, $\sr{86}$ \cite{Borkowski2014}, and in $\sr{84}$ \cite{Stellmer2012}.
In addition, two-color photoassociation in $\sr{88}$ was used to calculate the scattering lengths of all the strontium isotopes \cite{MartinezdeEscobar2008} and several subradiant $1_g$ states have been probed in $\sr{88}$ \cite{McGuyer2014}.
The ground \cite{Stein2008,Stein2010} and excited \cite{Stein2011} state molecular potentials have also been explored by Fourier transform spectroscopy.

In addition to probing the shapes of the molecular potentials, photoassociation efforts are motivated by interest in creating ground-state molecules.
Ground state molecules have been proposed as a platform for precision measurements to study the time-variation of the proton-electron mass ratio \cite{Zelevinsky2008, Kotochigova2009a}, and the fine-structure constant \cite{Beloy2011}. 
The production of ground state molecules has been demonstrated by decay from excited-molecular states in $\sr{88}$ \cite{Reinaudi2012} and by using stimulated Raman adiabatic passage (STIRAP) in $\sr{84}$ \cite{Stellmer2012,Ciamei2017,Ciamei2017a}. 
In particular, the technique in Ref.~\cite{Ciamei2017a} may offer a path towards creating a molecular Bose-Einstein condensate (BEC).

Another motivation for studying narrow-line photoassociation resonances stems from the prospect of using them as optical Feshbach resonances (OFRs) to tune the \textit{s}-wave scattering length of ground-state atoms \cite{Fedichev1996,Bohn1999,Zelevinsky2006,Nicholson2015a}. 
The application of OFRs to strontium and other alkaline-earth elements is particularly interesting because the spinless ground-state of the bosonic isotopes precludes the use of magnetic Feshbach resonances.
In addition, OFRs offer the possibility of controlling atomic interactions with increased temporal and spatial resolution compared to magnetic resonances. 
OFRs were first demonstrated in alkali atomic systems, but their utility has been limited by the fact that significant scattering-length modification is accompanied by very rapid atom loss \cite{Fatemi2000,Theis2004}.
Early theoretical work suggested that the atomic loss could be reduced by using the narrow intercombination transitions of alkaline-earth atoms \cite{Ciurylo2005}.
In particular, there was hope that this technique could be applied to $\sr{88}$, which is the most abundant isotope of strontium but whose nearly vanishing scattering length of $-1.4\, a_0$ \cite{MartinezdeEscobar2008}, where $a_0=0.0529$~nm is the Bohr radius, prevents evaporative cooling from being effective.
Two groups successfully used an OFR to modify the scattering length of $\sr{88}$; however these results were accompanied by rapid atom loss, limiting the experimental lifetime to the order of a few ms or less \cite{Blatt2011,Yan2013b}. 
Nevertheless, OFRs may still prove to be useful in other situations or systems. 
For example, OFRs were used to modify the scattering lengths of $^{176}$Yb and $^{172}$Yb by more than $200\, a_0$ with minimal loss \cite{Enomoto2008}. 
At low energies, the behavior of an OFR is characterized by a parameter known as the optical length, which can be determined by measuring the atomic loss rate as a function of photoassociation laser intensity \cite{Ciurylo2005}.
These line strengths have been measured for prominent resonances in $\sr{88}$ \cite{Zelevinsky2006,Blatt2011} and $\sr{86}$ \cite{Borkowski2014}, but not, prior to this work, for $\sr{84}$.

In Ref.~\cite{Borkowski2014}, Borkowski \etal\ adjusted \textit{ab initio} potentials from Ref.~\cite{Skomorowski2012} to reproduce the known photoassociation resonances of the bosonic strontium isotopes. 
They developed independent potentials for each isotope (see \cite{Borkowski2014} Sec.~III) that, although mostly identical, included a couple fitting parameters that were tuned to the spectrum of each isotope. 
These potentials replicated most of the known lines, but also pointed to some open questions. 
Though the binding energy spectrum of $\sr{88}$ was well measured with 11 known lines of both $0_u^+$ and $1_u$ symmetry in the Hund's case (c) representation, they only had four resonances each for $\sr{84}$ and $\sr{86}$. 
Their model only included one $\sr{86}$ $1_u$ resonance, did not consider any $\sr{84}$ lines with $1_u$ symmetry, and did not reproduce one of the observed $\sr{84}$ $0_u^+$ lines.

In addition, Borkowski \etal\ constructed a mass-scaled model for the three bosonic isotopes \cite{Borkowski2014}. 
A mass-scaled model consists of a single set of potentials that recreates the photoassociation spectra of all the bosonic isotopes of strontium simultaneously while only varying the molecular reduced mass of each isotope. 
Interestingly, in developing their model, Borkowski \etal\ discovered that they needed to include the $\term{1}{S}{0} + \term{1}{D}{2}$ $0_u^+$ $(^1\Sigma_u^+)$ potential, which has an avoided crossing with the $\term{1}{S}{0} + \term{3}{P}{1}$ $0_u^+$ $(^3\Pi_u)$ potential at short range. 
This perturbing state was measured by Stein \etal~\cite{Stein2011} and theoretically described by Skomorowski \etal~\cite{Skomorowski2012}. 
However, as Borkowski \etal\ pointed out, this mass-scaled, multi-channel model was insufficiently constrained given the known resonances in $\sr{84}$ and $\sr{86}$.

In this work, we clear up these issues by measuring additional resonances in $\sr{84}$ and $\sr{86}$ and by developing an updated theoretical model. 
The paper is organized as follows. 
First, we describe the general experimental procedure in Sec.~\ref{sec:exp}. 
Next, Sec.~\ref{sec:bindingEnergies} discusses the measurement of photoassociation resonances in $\sr{84}$ and $\sr{86}$ down to binding energies of $\simeq - 5$~GHz, bringing the total number of known resonances up to seven for both $\sr{84}$ and $\sr{86}$. 
In measuring the new lines, we discovered that some of the previously measured resonance positions could not be reproduced and measured new locations for those resonances. 
With these new spectra we build theoretical models that reproduces the known spectrum to better than 1~MHz in Sec.~\ref{sec:theory}. 
Our mass-scaled model reproduces the known binding energies for the $\sr{86}$ and $\sr{88}$ isotopes spanning many GHz and includes lines with both $0_u^+$ and $1_u$ symmetry. 
Due to the avoided crossing with the $\term{1}{S}{0} + \term{1}{D}{2}$ $^1\Sigma_u^+$ potential, the $\sr{84}$ spectrum is highly perturbed, and therefore we treat this isotope as a special case with independent quantum-defect parameters. 
In Sec.~\ref{sec:measureOFR}, we describe how the behavior of an OFR is characterized by the optical length and the measurement of these parameters for multiple $\sr{84}\ 0_u^+$ lines.
Finally, we conclude in Sec.~\ref{sec:conclusion} and discuss potential applications of this work.

\section{\label{sec:exp}Experimental Procedure}

\begin{figure*}

\includegraphics[width=\linewidth]{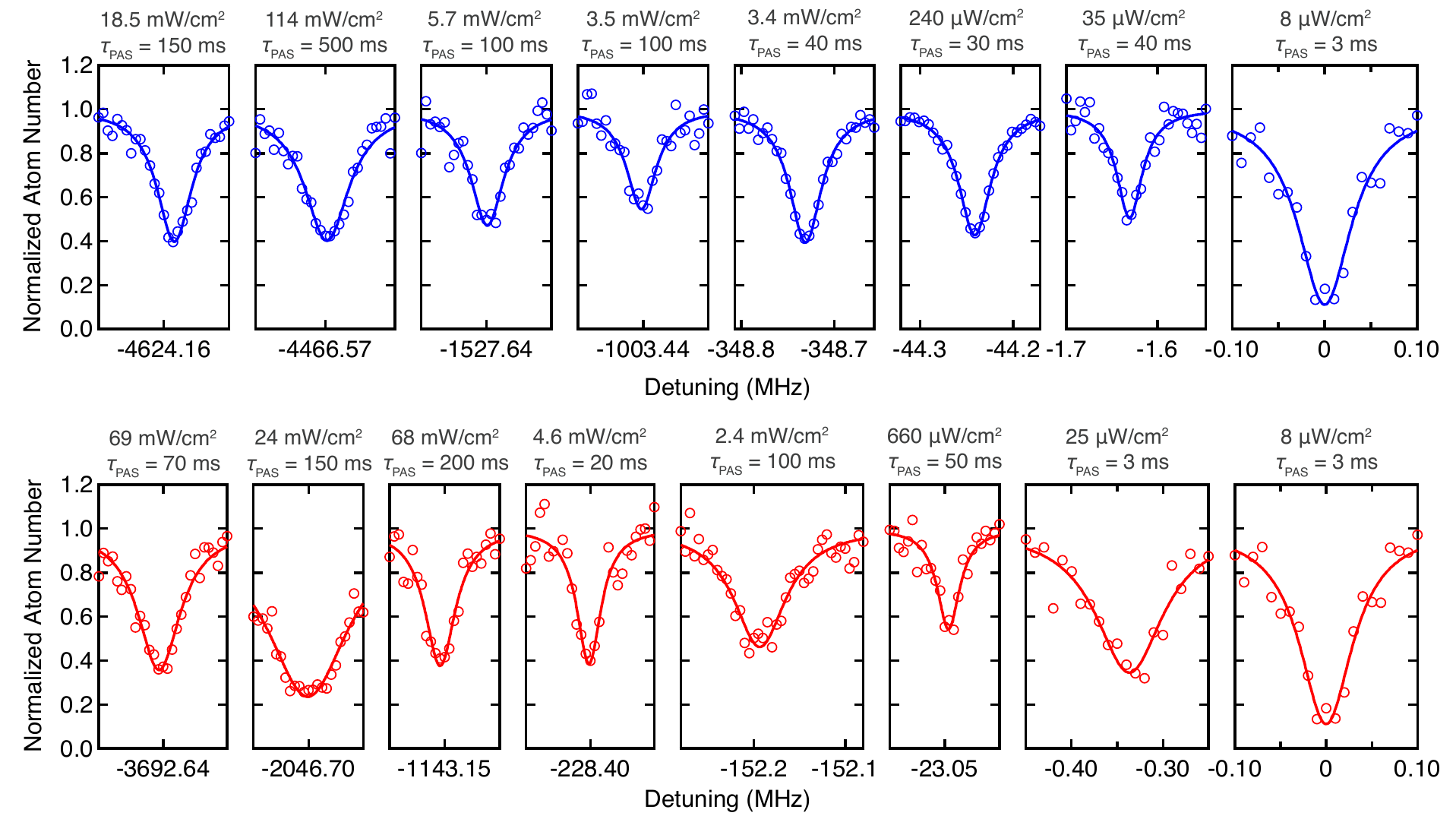}
\caption{\label{fig:centerScans}(Color online) Photoassociation resonances for all $^{86}$Sr (top, blue) and $^{84}$Sr (bottom, red) features measured as a function of detuning. The solid lines are Lorentzian fits to the data and the horizontal scale is consistent for all plots. The vertical scale of each scan is normalized to the expected atom number far from resonance extracted from the Lorentzian fit. The scans on the far right are the atomic $^1$S$_0$ $\rightarrow$ $^3$P$_1$ transition used to calibrate the absolute detuning. The photoassociation laser intensity and total pulse duration is indicated above each scan.}
\end{figure*}

Our cooling procedure is similar to those used in other ultracold strontium gas experiments \cite{Stellmer2014a,Borkowski2014,Stellmer2012}. 
For more details on our apparatus see Refs.~\cite{Barker2015,Barker2016}.
We use a Zeeman slower and 3D magneto-optical-trap (MOT) operating on the $\term{1}{S}{0}\rightarrow\term{1}{P}{1}$ transition to slow and capture atoms from a hot atomic beam. 
After a few seconds of loading, the atoms are cooled to $\approx$ 1~mK and transferred to a secondary MOT operating on the narrow ($\Gamma=2\pi\times7.5$\kHz) $\term{1}{S}{0}\rightarrow\term{3}{P}{1}$ transition at 689\nm.
We compress and cool the atoms in the secondary MOT to $\approx 1$ $\mu$K before loading them into an optical dipole trap (ODT) at 1064\nm. 

After loading into the dipole trap, our experimental procedure is slightly different depending on the isotope.
The large \textit{s}-wave scattering length of $823\, a_0$ for $\sr{86}$ \cite{MartinezdeEscobar2008} causes these samples to suffer from rapid three-body losses at large densities \cite{Ferrari2006,Stellmer2010}. 
Therefore, we perform photoassociation of $\sr{86}$ in a thermal gas as opposed to a BEC to maintain favorable signal-to-noise ratios from larger atom numbers.
We use a single-beam dipole trap consisting of a pancake-shaped beam propagating the horizontal plane with a vertical (horizontal) $1/e^2$ waist of 22.8~$\mu$m (228~$\mu$m). 
After evaporating for 1.0~s to a trap depth, $U/k_B$, of approximately 2.5~$\mu$K, we produce a sample consisting of $\simeq 10^6$ atoms at a temperature of about 150~nK and peak density of $\simeq 2 \times 10^{12}$~cm$^{-3}$.

The $\sr{84}$ isotope can be readily evaporated to degeneracy \cite{DeEscobar2009,Stellmer2009a,Stellmer2013a}, so we perform photoassociation of this isotope in a BEC.
We can then ignore thermal lineshape effects \cite{Ciurylo2004} and the large density enhances the photoassociation signal.
Our dipole trap for $\sr{84}$ includes a vertical beam with a $1/e^2$ waist of 72\um\ in addition to the horizontal beam mentioned above. 
After approximately two seconds of evaporation, we create a nearly pure Bose-Einstein condensate with approximately $10^5$ atoms. 
Typical trap frequencies after evaporation are $\lbrace \omega_x,\omega_y, \omega_z \rbrace = 2\pi \times \lbrace 40,40,140 \rbrace$ Hz and the BEC has a  chemical potential, $\mu/h$, of about 1~kHz.
The peak density is about $10^{14}~\mathrm{cm}^{-3}$ and the final trap depth is $\simeq 2.5~\mu$K.

The photoassociation laser is referenced to a master laser operating near the 689~nm intercombination transition. 
We stabilize the master with a Pound-Drever-Hall lock to a very high finesse ($\mathcal{F}>200\,000$) cavity made from Ultra-Low-Expansion (ULE) glass.
The cavity is operated in a temperature stabilized vacuum chamber in order to minimize thermal and pressure drifts. 
The linewidth of the master laser is $\leq 200$~Hz, as determined by measuring the joint linewidth of the master with an independent laser locked to a separate, similar cavity. 
We measured the long term drift of the cavity to be $\simeq 28$~mHz/s by monitoring the $\term{1}{S}{0} \rightarrow \term{3}{P}{1}$ transition frequency over several months.
The photoassociation laser is stabilized relative to the master by an optical phase-locked loop (OPLL) \cite{Appel2008}. 
This locking method gives us a great deal of flexibility, since we can vary the detuning of the PAS laser with respect to the atomic resonance over many GHz simply by changing the reference frequency supplied to the OPLL. 
The PAS laser light is delivered to the experiment by a single-mode optical fiber and has a $1/e^2$ waist of 1.63~mm at the location of the atomic sample.
We stabilize the PAS intensity via feedback to an acousto-optic modulator (AOM).
Frequency scans were either performed at zero magnetic field or with a small bias field of $20~\mu$T (200~mG) parallel to the polarization of the photoassociation laser.

In order to avoid AC Stark shifts from the 1064~nm trapping beams, we turn on and off the dipole traps with a 50~\% duty cycle and period of 500~$\mu$s so that we can apply the PAS laser while the 1064~nm beams are off. 
We vary the total amount of time the PAS laser is on from 10~ms to 500~ms in order to limit the maximum photoassociation atom number loss to $\approx$~50~\% for the intensity and resonance under investigation.
For the results in Sec.~\ref{sec:bindingEnergies}, we use a single  intensity for each resonance as indicated in Fig.~\ref{fig:centerScans}, whereas for the optical length measurements in Sec.~\ref{sec:measureOFR}, we repeat the procedure with five different intensities for each resonance.
After applying the PAS laser, we measure the remaining number of atoms by absorption imaging using a 10~$\mu$s pulse from a laser beam resonant with the $\term{1}{S}{0} \rightarrow \term{1}{P}{1}$ transition.
To avoid systematic errors we limit the maximum optical depths of our $\sr{84}$ BEC samples to $\leq 2.0$ by allowing the condensates to expand for 25~ms before imaging. 
We image our thermal $\sr{86}$ samples after an expansion time of 12~ms.

\section{\label{sec:bindingEnergies}Measured Binding Energy Spectrum}

\subsection{\label{bindingEnergyMethod}Method and Uncertainties}
We measured the binding energies of seven photoassociation resonances each for $\sr{84}$ and $\sr{86}$.
To determine these binding energies, we monitor atom number as the detuning of the PAS laser is varied. 
In the vicinity of a $\term{1}{S}{0}+\term{3}{P}{1}$ molecular bound state, the PAS laser forms molecules, which leads to atom loss as the excited dimers quickly decay into ground state molecules or free atoms with enough kinetic energy to escape our shallow dipole trap. 
An example atomic loss feature for each of the resonances we measured in $\sr{84}$ and $\sr{64}$ is shown in Fig.~\ref{fig:centerScans}.

We fit the atomic loss features as a function of detuning to extract the binding energy, $E_b$, of each resonance.
Since our thermal $\sr{86}$ samples have temperatures below the atomic recoil temperature, $T_R \simeq 460$~nK, Doppler and thermal shifts cause broadening and shifts of the atomic line feature on the order of several kHz.
We account for these effects by using the approach described in the appendix of \cite{Borkowski2014}, which is based on the lineshape formalism of \cite{Ciurylo2004}.
These thermal effects are negligible in a BEC, so we extract the resonance position of our $\sr{84}$ samples by fitting the photoassociation loss features to a simple Lorentzian lineshape.
We also take into account a shift due to photon recoil of $E_\mathrm{rec}/h =  \hbar^2 k_\mathrm{las}^2 / 4 h m \simeq 2.5$~kHz, where $k_\mathrm{las}$ is the wavenumber of the photoassociation laser and $m$ is the mass of a single strontium atom, for both BEC and thermal samples.
Each scan is repeated at least three times and the results are averaged.

In order to assign uncertainties to the measured binding energies, we quantified a number of sources of error.
Though the long-term drift of our photoassociation laser is very small
(28~mHz/s), we also observe intermediate timescale drifts on the order of tens of kHz over a few hours, likely due to temperature fluctuations in the lab. 
In order to minimize errors due to these drifts of the photoassociation laser, we calibrate the absolute frequency by scanning over the atomic resonance before and after each scan of a photoassociation resonance. 
The typical laser drift is $\simeq 3$~kHz during these scans, which is similar to the typical statistical uncertainty associated with extracting the resonance position from the thermal lineshape or Lorentzian fits ($\simeq 2$~kHz). 
AC Stark shifts due to the ODT are eliminated by turning off the 1064~nm trapping beams while applying the PAS beam, as described in Sec.~\ref{sec:exp}.

We did not perform a systematic investigation into the mean field shifts for our system. 
However, Stellmer \textit{et al.} \cite{Stellmer2012} measured mean field shifts for the $\sr{84}$ $0_u^+, \nu = -3$ transition to be about 2.4(5)~mHz/atom. 
For our BECs with atom number $\approx 10^5$, this would correspond to shifts of about 250 Hz. 
We do not observe any dependence of the binding energies on the intensity of the PAS laser for the relatively low intensities used in this work.
We calculate AC Stark shifts due to the photoassociation laser to be $< 500$~Hz.
Taking all these error sources into account, we estimate the uncertainty in our binding energies to be 10~kHz, which is dominated by laser drift and the statistical uncertainties of our fits.


\subsection{\label{bindingEnergyResults}Results and Discussion}

\begin{table*}
\centering
\caption[Photoassociation Resonances for Sr]{\label{tab:PASresults}Experimentally measured and numerically calculated photoassociation resonances for the bosonic isotopes of strontium. The $\nu$ index counts the vibrational levels down from the dissociation limit. For $^{86}$Sr and $^{84}$Sr, we measured several new resonances, confirmed some previous measurements, and corrected several lines observed by other experiments that are likely to be spurious. Since the $^{88}$Sr binding energies are well known, we did not perform additional measurements on that isotope and instead used the results of Ref.~\cite{Zelevinsky2006} to inform our mass-scaled model. The last five columns are based on the numerical calculations described in Sec.~\ref{sec:theory}. The mass-scaled model describes the resonances of $^{86}$Sr and $^{88}$Sr and a special case of the same model with a different set of quantum defect parameters describes the resonances for $^{84}$Sr. The seventh column shows the error $\delta = E^\text{th}/h-E^\text{exp}/h$ between each theoretical energy and the corresponding measurement. The last three columns give the projections of the eigenstates onto each channel, labeled by P$(J_a,\Omega)$. The symmetry label assigned in the second column is based upon the largest of these projections for each line. Due to the proximity of the $^{86}$Sr lines at $-4624$~MHz and $-4467$~MHz, there is significant Coriolis mixing between these two states.} 
\begin{ruledtabular}
\begin{tabular}{c c c d{7} d{7} d{6} d{4} d{3} d{3} d{3}}

 & &  & \multicolumn{3}{c}{Binding Energy $E_b/h$ (MHz)} & & & & \\[0.5ex]
 \cline{4-6}  \T

Isotope & Symmetry & $\nu$ & \multicolumn{1}{c}{Expt.\ (this work)} & \multicolumn{1}{c}{Expt. \cite{Zelevinsky2006}} & \multicolumn{1}{c}{Theory (this work)} & \multicolumn{1}{c}{$\delta$ (MHz)} & \multicolumn{1}{c}{P(1,0)} &\multicolumn{1}{c}{P(1,1)}& \multicolumn{1}{c}{P(2,1)}\\[0.5ex]
\hline \T
$\sr{88}$ & $0_u^+$ & $-1$ &  & -0.435(37) & -0.406 & 0.029 & 0.917 & 0.083 & 0.000\\[0.5ex]
 & $0_u^+$ & $-2$ &  & -23.932(33) & -23.871 & 0.061 & 0.988 & 0.011 & 0.000 \\[0.5ex]
 & $0_u^+$ & $-3$ & & -222.161(35) & -222.104 & 0.057 & 0.995 & 0.005 & 0.000 \\[0.5ex]
 & $0_u^+$ & $-4$ & & -1084.093(33) & -1083.862 & 0.231 & 0.996 & 0.004 & 0.000 \\[0.5ex]
 & $0_u^+$ & $-5$ & & -3463.280(33) & -3463.064 & 0.216 & 0.992 & 0.008 & 0.000 \\[0.5ex]
 & $0_u^+$ & $-6$ &  & -8429.650(42) & -8429.805 & -0.155 & 0.840 & 0.159 & 0.000 \\[0.5ex]
 & $1_u$ & $-1$ & & -353.236(35) & -353.682 & -0.446 & 0.005 & 0.995 & 0.000\\[0.5ex]
 & $1_u$ & $-2$ & & -2683.722(32) & -2684.181 & -0.459 & 0.008 & 0.991 & 0.001  \\[0.5ex]
  & $1_u$ & $-3$ &  & -8200.163(39) & -8200.222 & -0.059 & 0.160 & 0.838 & 0.002  \\[0.5ex]
 
Isotope  & Symmetry & $\nu$ & \multicolumn{1}{c}{Expt.\ (this work)} & \multicolumn{1}{c}{Expt. \cite{Borkowski2014}} & \multicolumn{1}{c}{Theory (this work)} & \multicolumn{1}{c}{$\delta$ (MHz)} & \multicolumn{1}{c}{P(1,0)} &\multicolumn{1}{c}{P(1,1)}& \multicolumn{1}{c}{P(2,1)}\\[0.5ex]
\hline \T
$\sr{86}$ & $0_u^+$ & $-1$ & -1.625(10) & -1.633(10) & -1.560 & 0.065 & 0.949 & 0.051 & 0.000\\[0.5ex]
 & $0_u^+$ & $-2$ & -44.233(10) & -44.246(10) & -44.141 & 0.092 & 0.992 & 0.008 & 0.000 \\[0.5ex]
 
 & $0_u^+$ & $-3$ & -348.729(10)&-348.742(10) & -348.798 & -0.069 & 0.994 & 0.006 & 0.000 \\[0.5ex]
 & $0_u^+$ & $-4$ & -1527.645(10) & & -1527.934 & -0.289 & 0.989 & 0.011 & 0.000 \\[0.5ex]
 & $0_u^+$ & $-5$ & -4466.572(10) & & -4467.018 & -0.446 & 0.650 & 0.349 & 0.001 \\[0.5ex]
 & $0_u^+$ & $-6$ &  & & -10448.722 &  & 0.996 & 0.004 & 0.000 \\[0.5ex]
 & $1_u$ & $-1$ & -1003.449(10) & -159.984(50) & -1003.194 & 0.255 & 0.011 & 0.988 & 0.001\\[0.5ex]
 & $1_u$ & $-2$ & -4624.155(10) & & -4623.268 & 0.887 & 0.350 & 0.649 & 0.001  \\[0.5ex]
  & $1_u$ & $-3$ &  & & -11979.275 &  & 0.004 & 0.993 & 0.003  \\[0.5ex]

Isotope  & Symmetry & $\nu$ & \multicolumn{1}{c}{Expt.\ (this work)} & \multicolumn{1}{c}{Expt. \cite{Stellmer2012}} & \multicolumn{1}{c}{Theory (this work)} & \multicolumn{1}{c}{$\delta$ (MHz)} & \multicolumn{1}{c}{P(1,0)} &\multicolumn{1}{c}{P(1,1)}& \multicolumn{1}{c}{P(2,1)}\\[0.5ex]
\hline \T
$\sr{84}$ & $0_u^+$ & $-1$ & -0.338(10) & -0.32(1) & -0.257 & 0.081 & 0.902 & 0.098 & 0.000 \\[0.5ex]
 & $0_u^+$ & $-2$ & -23.050(10) & -23.01(1) & -22.983 & 0.067 & 0.987 & 0.013 & 0.000\\[0.5ex]
 & $0_u^+$ & $-3$ & -228.406(10) & -228.38(1) & -228.827 & -0.421 & 0.984 & 0.016 & 0.000 \\[0.5ex]
 & $0_u^+$ & $-4$ & -1143.161(10) & -1288.29(1) & -1143.539 & -0.378 & 0.998 & 0.002 & 0.000\\[0.5ex]
 & $0_u^+$ & $-5$ & -3692.645(10) & & -3692.448 & 0.197 & 0.999 & 0.001 & 0.000 \\[0.5ex]
  & $0_u^+$ & $-6$ & & & -9000.576 & & 0.999 & 0.001 & 0.000 \\[0.5ex]
 & $1_u$ & $-1$ & -152.193(10) & -351.45(2) & -152.421 & -0.228 & 0.015 & 0.984 & 0.000\\[0.5ex]
 & $1_u$ & $-2$ & -2046.703(10) & & -2046.640 & 0.063 & 0.001 & 0.997 & 0.001 \\[0.5ex]
 & $1_u$ & $-3$ & & & -7058.199 & & 0.002 & 0.996 & 0.002 \\[0.5ex]
\end{tabular}

\end{ruledtabular}
\end{table*}

The measured photoassociation resonances are shown in Table \ref{tab:PASresults}.
The measured binding energies of the three least bound $0_u^+$ levels for both $\sr{86}$ and $\sr{84}$ are in agreement with previous experimental measurements. 
However, we could not reproduce the previously measured  lines at $-160$~MHz ($\sr{86}$ $1_u$ $\nu=-1$) \cite{Borkowski2014}, $-1288$~MHz ($\sr{84}$ $0_u^+$ $\nu=-4$), and $-351$~MHz ($\sr{84}$ $1_u$ $\nu=-1$) \cite{Stellmer2012}. 
We cannot say with certainty that the previous results were spurious, but we observe that given the very small saturation intensity of the $\term{1}{S}{0} \rightarrow \term{3}{P}{1}$ atomic transition ($I_\mathrm{sat} = 3~\mu\mathrm{W/cm}^2$), a small amount of sideband noise on the photoassociation laser that happens to be nearly resonant with the atomic transition can cause losses that are easily confused as a photoassociation resonance. 
For instance, our PAS beam originates from the $-1$ diffracted order of a single-passed AOM operating at a frequency of $\Omega_\mathrm{AOM}$.
When scanning the detuning of our PAS beam, we observe dips at detunings of $-\Omega_\mathrm{AOM}$ and $-2 \Omega_\mathrm{AOM}$.
We attribute the first dip to a small amount of non-diffracted light from the AOM being collected by the optical fiber, which is then resonant with the atomic transition at a detuning of $-\Omega_\mathrm{AOM}$.
We believe that the second dip can similarly be attributed to the small back-reflection of acoustic waves in the AOM crystal adding a small component of light whose frequency is upshifted, instead of downshifted, by the AOM and is therefore resonant with the atomic transition at a detuning of $-2\Omega_\mathrm{AOM}$.
With the potential for extremely small sidebands to cause significant atomic loss, it can be difficult to identify and eliminate spurious signals.

By measuring multiple $0_u^+$ and $1_u$ resonances for $\sr{86}$ and $\sr{84}$, we can verify that the relative spacings are consistent with expectations, giving us more confidence in our results. 
In addition, our measurement of the $\sr{84}$ $0_u^+,~\nu=-4$ line at $-1143$~MHz instead of $-1288$~MHz from Ref.~\cite{Stellmer2012} is in much better agreement with the theoretical models of Ref.~\cite{Borkowski2014}.
We have measured new $\sr{86}$ and $\sr{84}$ resonances, extending to binding energies of nearly $-5$~GHz.
Next, we will extend the theoretical approach of Ref.~\cite{Borkowski2014} in order to account for the updated photoassociation spectrum.

\section{\label{sec:theory}Theoretical Model}

Using the experimental results from Sec.~\ref{sec:bindingEnergies}, we can provide an updated mass-scaled model of photoassociation in Sr. Ref.~\cite{Borkowski2014} created a mass-scaled model that reproduces the $0_u^+$ series, and we extend this work to include the $1_u$ series. However, our mass-scaled model does not include $\sr{84}$. Ref.~\cite{Borkowski2014} indicates that resonances in $\sr{84}$ are strongly perturbed by the $\term{1}{S}{0} + \term{1}{D}{2}$ $^1\Sigma_u^+$ state. Hence, a mass-scaled model for all isotopes would require a careful fine-tuning of the perturbing potential, and we do not include this perturbing state in our model. Instead, we consider $\sr{84}$ as a special case, where our model for this isotope only differs from our mass-scaled model by the values of its quantum defect fitting parameters. These models accurately reproduce all of the measured resonance energies reported in Table~\ref{tab:PASresults}.

We construct our mass-scaled model using the notation of Ref.~\cite{Borkowski2014}. The molecular Hamiltonian has the form,
\begin{equation}
\label{eq:Hamiltonian}
 H = T + V_\text{int} + V_\text{rot}.
\end{equation}
The first term $T=-(\hbar^2/2\mu)\mathrm{d}^2/\mathrm{d}r^2$ is the kinetic energy operator, where $\mu=m/2$ is the molecular reduced mass and $r$ is the internuclear separation. The second term $V_\text{int}$ is the molecular interaction potential, and $V_\text{rot}$ is the rotational energy of the molecule, including Coriolis couplings. We determine photoassociation energies by numerically solving the Schr\"{o}dinger equation $H\Psi=E\Psi$, using the matrix DVR method \cite{Tiesinga1998} with non-linear coordinate scaling. 

Our model uses the Hund's case (a) potentials $\mathcal{V}(^3\Pi_u,r)$ and $\mathcal{V}(^3\Sigma_u^+,r)$ from Ref.~\cite{Skomorowski2012}, and we use these potentials to construct a three-channel model in Hund's case (c) \cite{Borkowski2014,Mies1978},
\begin{widetext}
\begin{equation}
 V_\text{int}=
 \begin{pmatrix}
 \ \mathcal{V}(^3\Pi_u,r) - C_{3,0}/r^3 & 0 & 0\\
 \ 0 & \frac{1}{2}\left(\mathcal{V}(^3\Sigma_u^+,r) + \mathcal{V}(^3\Pi_u,r)\right) - C_{3,1}/r^3 & \frac{1}{2}\left(\mathcal{V}(^3\Sigma_u^+,r) - \mathcal{V}(^3\Pi_u,r)\right) \\
 \ 0 & \frac{1}{2}\left(\mathcal{V}(^3\Sigma_u^+,r) - \mathcal{V}(^3\Pi_u,r)\right) & \frac{1}{2}\left(\mathcal{V}(^3\Sigma_u^+,r) + \mathcal{V}(^3\Pi_u,r)\right) + \xi \\
 \end{pmatrix}.
\end{equation}
\end{widetext}
All of these channels have $J=1$ and $S=1$, where $J$ is the total angular momentum and $S$ is the electronic spin angular momentum. Each channel is associated with a different value of $(J_a,\Omega)$, where $J_a$ is the total electronic angular momentum and $\Omega$ is the projection of $J_a$ onto the internuclear axis. The channel order is $(J_a,\Omega)=(1,0)$, (1,1), and (2,1). The coefficients $C_{3,0}=0.0152356615~E_h a_0^3$ and $C_{3,1}=-C_{3,0}/2$ represent the resonant dipole interaction for the (1,0) and (1,1) channels, respectively \cite{Borkowski2014}, and $E_h=4.36 \times 10^{-18}$~J. The parameter $\xi= h c (387.358$~cm$^{-1})$ is the atomic spin-orbit constant \cite{NIST_ASD}, and the rotational energy is,
\begin{equation}
 V_\text{rot}=\frac{\hbar^2}{2\mu r^2}
 \begin{pmatrix}
  \ 4 & -2\sqrt{2} & 0 \\
  \ -2\sqrt{2} & 2 & 0 \\
  \ 0 & 0 & 6 \\
 \end{pmatrix}.
\end{equation}

\begin{figure}[h]
\includegraphics[width=.99\columnwidth]{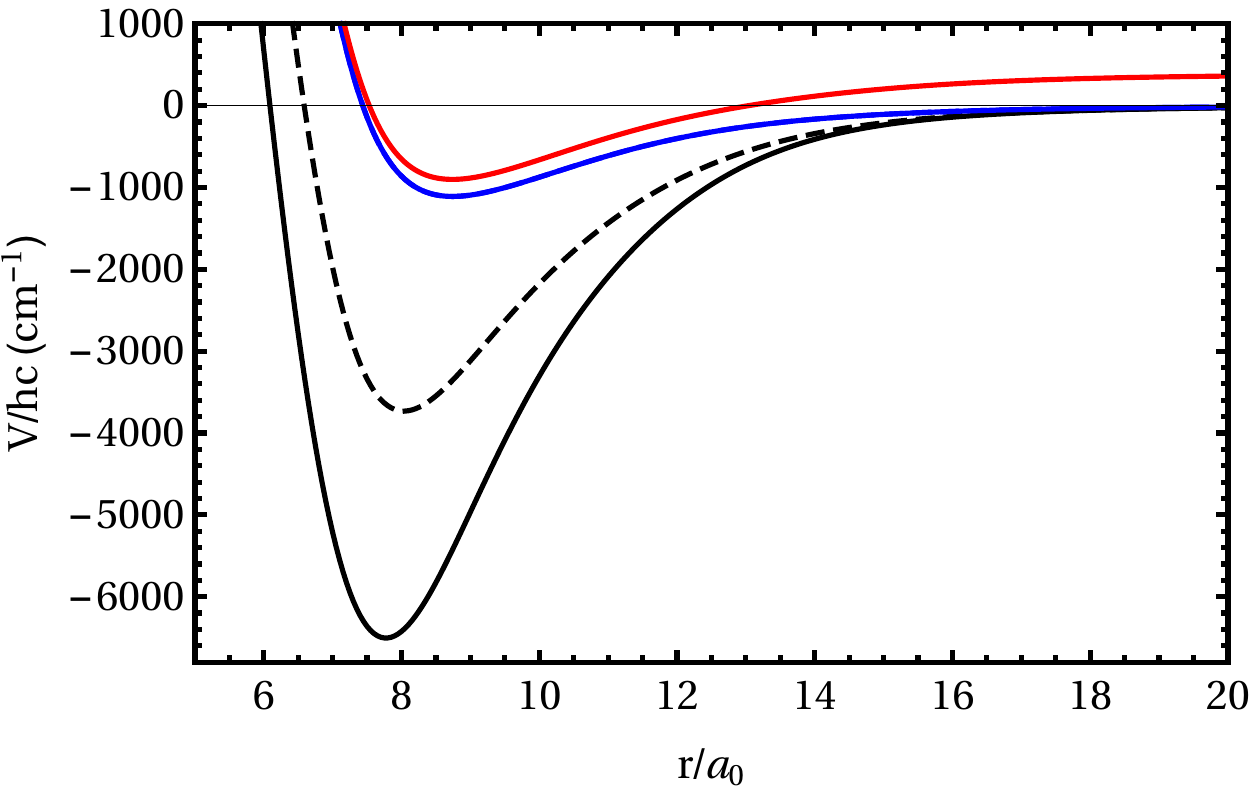}
\caption[Adiabatic Potentials]{\label{fig:adiabats}(Color online) The adiabatic potentials of the two-channel model (dashed curves) and the three-channel model (solid curves) relative to the $\term{1}{S}{0} + \term{3}{P}{1}$ limit. The adiabatic $1_u$ potential (black) is much deeper in the three-channel model. The adiabatic $0_u^+$ potential (blue) is nearly the same in both models, and the two curves are indistinguishable on this scale. The third adiabatic potential (red),  associated with the ($J_a=2$, $\Omega=1$) channel, dissociates to the $\term{1}{S}{0} + \term{3}{P}{2}$ limit and is shifted up by the spin-orbit constant~$\xi$.}
\end{figure}

Ref.~\cite{Borkowski2014} uses only the first two channels of this model, which we refer to as the two-channel model. Fig.~\ref{fig:adiabats} shows the adiabatic potentials, $V=V_\mathrm{int}+V_\mathrm{rot}$, for both the two-channel model and the full, three-channel model. In the two-channel model, the adiabatic $1_u$ potential is unphysically shallow ($\min\left[V/hc\right] \approx 3500$~cm$^{-1}$) and does not support enough bound states to produce the correct mass-scaling behavior. Adding the third channel creates a drastic increase in the depth of the adiabatic $1_u$ potential. Including this channel ensures that the adiabatic $1_u$ potential closely resemble the $^3\Sigma_u^+$ potential at short range, with a depth of approximately 6500~cm$^{-1}$. This three-channel model will be our starting point to achieve an accurate mass-scaling of the $1_u$ series.

The starting point for our fitting process are the potential parameters from Ref.~\cite{Borkowski2014} that were generated using the two channel model. Adding the third channel shifts the resonance spectrum of both the $0_u^+$ and the $1_u$ series. To correct for this shift, we fit our three-channel, mass-scaled model to the measured resonance energies for $\sr{86}$ (this work) and $\sr{88}$ reported in Table~\ref{tab:PASresults}. We perform this fit by adjusting the quantum defect parameters $\alpha(^3\Sigma_u^+)$ and $\alpha(^3\Pi_u)$ of the Hund's case (a) potentials (as defined in Ref.~\cite{Borkowski2014}) to minimize the rms error between the energies produced by our theoretical model $E^\text{th}$ and the experimentally measured energies $E^\text{exp}$. We define the rms error to be,
\begin{equation}
 \delta_\text{rms} = \frac{1}{h}\sqrt{\frac{1}{N}\sum_{i=1}^N \left(E_i^\text{th}-E_i^\text{exp}\right)^2},
\end{equation}
where $N$ is the number of measured resonances included in the fit.

In order to improve the accuracy of our model to $\delta_\text{rms}<1$~MHz, we find it necessary to modify the shape of our potential at long range. Table~\ref{tab:PASresults} includes energies as deep as $\approx -5$~GHz, and we find that the deeper energies show larger discrepancies between theory and experiment, which is consistent with Ref.~\cite{Borkowski2014}.
The long-range behavior is determined by the $C_6$ and $C_8$ dispersion coefficients, which are included in the definitions of $\mathcal{V}(^3\Pi_u,r)$ and $\mathcal{V}(^3\Sigma_u^+,r)$ \cite{Borkowski2014}.
We are able to reduce $\delta_\text{rms}$ by refitting the value of $C_8$, while keeping $C_6$ constant. This procedure modifies the shape of the long-range potential at shorter range, where the resonance energies become deeper, without changing the spacing between the most weakly bound levels. We find it sufficient to only change the value of $C_8$ for the $^3\Sigma_u^+$ state, which primarily affects the shape of the adiabatic $1_u$ potential.
Our fit minimizes the value of $\delta_\text{rms}$ as a function of the three parameters $C_8(^3\Sigma_u^+)$, $\alpha(^3\Sigma_u^+)$, and $\alpha(^3\Pi_u)$. Their optimal values are $C_8(^3\Sigma_u^+)=2.0237129\times 10^6~E_ha_0^8$, $\alpha(^3\Sigma_u^+)=0.064074850$, and $\alpha(^3\Pi_u)=1.973579586$, with $\delta_\text{rms}=0.326$~MHz.

We report the results of our mass-scaled model in Table~\ref{tab:PASresults}. The symmetry assignments in Table~\ref{tab:PASresults} are based on which channel has the largest projection for each state. Our model shows that the $\sr{86}$ resonances at $-4624$ and $-4467$~MHz have significant Coriolis mixing due to their proximity. Likewise, we use this optimal value of $C_8(^3\Sigma_u^+)$ in our isotope-specific model of $\sr{84}$. In this case, we determine the optimal values of the quantum defect parameters to be $\alpha(^3\Sigma_u^+)=0.064074227$ and $\alpha(^3\Pi_u) =1.975561082$ with $\delta_\text{rms}=0.247$~MHz, and we report the results for $\sr{84}$ in Table~\ref{tab:PASresults} as well.  

Combining the results for all three isotopes produces $\delta_\text{rms}=0.304$~MHz. This error is larger than the experimental uncertainties but is comparable to the error in the isotope-specific models of Ref.~\cite{Borkowski2014}. This level of accuracy can aid future experimental measurements of photoassociation resonances in Sr, including resonance energies as deep as approximately $-5$~GHz and those found in other isotopes, like $\sr{87}$. The inability of the theoretical results to more closely approach the experimental error bars is likely due to fine details of the true molecular potentials that are not captured by the model.

\section{\label{sec:measureOFR}Measuring Optical Lengths}

Having mapped out additional lines, we explore whether the prominent $\sr{84}$ lines could be useful as OFRs by measuring their optical lengths. 

\subsection{\label{sec:isolatedResTheory}Isolated Resonance Theory}

The isolated resonance model \cite{Bohn1999} is a good description of the behavior of optical Feshbach resonances as long as the detuning is small compared to the distance between the resonance and adjacent bound states \cite{Blatt2011,Nicholson2015a}. 
In this model \cite{Ciurylo2005,Ciurylo2006,Nicholson2015a}, the scattering length is given by $a = a_\mathrm{bg} + a_\mathrm{opt}$ where the background value, $a_\mathrm{bg}$, is modified by
\begin{equation}
a_{\mathrm{opt}} = \frac{\ell_\mathrm{opt} \Gamma_\mathrm{mol} \Delta}{\Delta^2 + (\eta \Gamma_\mathrm{mol})^2/4},
\end{equation}
and the two-body loss rate $K_2$ is given by
\begin{equation}
\label{eq:K2}
K_2 = \frac{4 \pi \hbar}{\mu \alpha_{deg}}\ \frac{\eta \Gamma_\mathrm{mol}^2 \ell_\mathrm{opt}}{\Delta^2 + (\eta \Gamma_\mathrm{mol} + \Gamma_{\mathrm{stim}})^2/4}.
\end{equation}
In these equations $\Gamma_\mathrm{mol}$, the molecular linewidth, is twice the atomic linewidth or $\Gamma_\mathrm{mol} = 2 \pi \times 15$~kHz. 
The parameter $\eta \geq 1$ is phenomenological and accounts for extra broadening typically measured in photoassociation experiments \cite{Theis2004,Zelevinsky2006,Blatt2011,Yan2013b,Borkowski2014,Kim2016} and the PAS laser detuning with respect to the photoassociation resonance is given by $\Delta$.
Here, the stimulated linewidth is given by $\Gamma_{\mathrm{stim}} = 2 k \ell_\mathrm{opt} \Gamma_\mathrm{mol}$ where $k$ is the wavenumber for the BEC or thermal sample. 
For a BEC, $k = \sqrt{21/8}/(2 R_\mathrm{TF})$, where $R_\mathrm{TF}$ is the Thomas-Fermi radius \cite{Yan2013b}.
For a thermal gas, $k = \sqrt{2 \mu E}/\hbar$, where $E$ is the kinetic energy of the colliding atoms.
The factor $\alpha_{deg} = 2$ (1) for BEC (thermal) samples accounts for the reduction of the inelastic scattering of a BEC compared to that of a thermal gas \cite{Jones2006, Ciurylo2004, Stoof1989}. 
Finally, $\lopt$, is a parameter known as the optical length.

From these equations, we see that the optical length is an important parameter to describe the effects of an OFR. 
The maximum change in scattering length is $a_\mathrm{opt} = \pm \ell_\mathrm{opt}$ when $\Delta = \pm \Gamma_\mathrm{mol}/2$ (and assuming $\eta = 1$).
The optical length is a measure of the coupling between the ground, scattering state and the excited, bound state and is defined as
\begin{equation}
\label{eq:optLength}
\ell_\mathrm{opt} = \frac{\lambda^3}{16 \pi c} \frac{f_{FC}}{k}f_{\mathrm{rot}} I,
\end{equation}
where $\lambda = 689.45$ nm is the wavelength of the atomic transition, the rotational factor $f_{\mathrm{rot}} = 1$ for $0_u^+$ resonances and $f_{\mathrm{rot}} = 2$ for $1_u$ resonances, $I$ is the intensity of the PAS laser, and $f_{FC}$ is the free-bound Franck-Condon factor per unit energy \cite{Nicholson2015a}. 
The Franck-Condon factor is a measure of the overlap between the ground and excited molecular wavefunctions and is defined as
\begin{equation}
\label{eq:franckCondonFac}
f_{FC} = \abs{\, \int_0^\infty \phi_e(r) \phi_g(E,r) dr \,}^2,
\end{equation}
where $\phi_e(r)$ is the excited bound molecular state wavefunction and $\phi_g(E,r)$ is the energy-normalized ground state scattering wavefunction.
The energy-normalized wavefunction has the long-range form of 
\begin{equation}
\label{eq:energyNormalizedWF}
\phi_g(E,r) \xrightarrow{r \rightarrow \infty} \sqrt{\frac{2 \mu}{\pi \hbar^2 k}} \sin[k(r-a_\mathrm{bg})]
\end{equation}
for \textit{s}-wave collisions at low energies \cite{Ciurylo2005,Ciurylo2006}.
Note that $f_{FC} \propto k$ for small $k$ \cite{Nicholson2015a}, so $\ell_\mathrm{opt}$ is independent of temperature.
The optical length is directly proportional to the photoassociation laser intensity, so the quantity $\ell_\mathrm{opt}/I$ is a constant parameter for each resonance.
We have measured the $\ell_\mathrm{opt}/I$ and $\eta$ broadening factors for the $\sr{84}$ $0_u^+$ states in order to characterize their properties.

\subsection{\label{sec:loptMeas}Method and Results}

The $\ell_\mathrm{opt}/I$ and $\eta$ parameters for the $\sr{84}$ $0_u^+$ resonances are measured with a procedure similar to that used in Ref.~\cite{Kim2016}.
We performed photoassociation spectroscopy of four $\sr{84}$ resonances using the method described in Sec.~\ref{sec:exp} for five different laser intensities. 
The measurements were performed in a nearly pure BEC.
At each intensity, we repeated the scan across the photoassociation resonance 4 to 6 times and averaged the results.

The photoassociation loss is modeled as ${\dot{n} (\mathbf{r})=-K_2n(\mathbf{r})^2}$ where $n(\mathbf{r})$ is the position-dependent atomic density. 
For a BEC in the Thomas-Fermi approximation, the total atom number changes according to \cite{Soding1999}
\begin{equation}
\label{eq:PASlossBEC}
\frac{d}{dt}\ln N =  -C_2 K_2 N^{2/5},
\end{equation} 
where 
\begin{equation}
\label{eq:PASC2}
C_2 = \frac{15^{2/5}}{14 \pi} \left( \frac{m \bar{\omega}}{\hbar a_\mathrm{bg}^{1/2}}\right)^{6/5},
\end{equation}
and $\bar{\omega}$ is the geometric mean of the harmonic trap frequencies.
The analytic solution to Eq.~(\ref{eq:PASlossBEC}) is
\begin{equation}
\label{eq:PASNumBEC}
N(\tau_\mathrm{PAS})= \frac{N_0}{\left(1 + \frac{2}{5} \tau_\mathrm{PAS} N_0^{2/5} C_2 K_2\right)^{5/2} },
\end{equation}
with initial atom number $N_0$ and total PAS laser pulse time $\tau_\mathrm{PAS}$ \cite{Kim2016}.
By combining Eqs.~(\ref{eq:PASNumBEC}), (\ref{eq:PASC2}), and (\ref{eq:K2}), we can fit the loss features to extract $\ell_\mathrm{opt}$ and $\eta$ for each resonance and intensity. 
Since $\Gamma_\mathrm{stim} \ll \Gamma_\mathrm{mol}$ for the intensities used here, we ignore $\Gamma_\mathrm{stim}$ in the denominator of Eq.~(\ref{eq:K2}). 
We confirmed that atom losses due to other processes, such as one-body loss from collisions with background gas molecules or far off-resonant scattering from the atomic transition, are negligible for the parameters used in this study.

\begin{figure}

\begin{subfigure}{0.235\textwidth}

\includegraphics[width=1\linewidth]{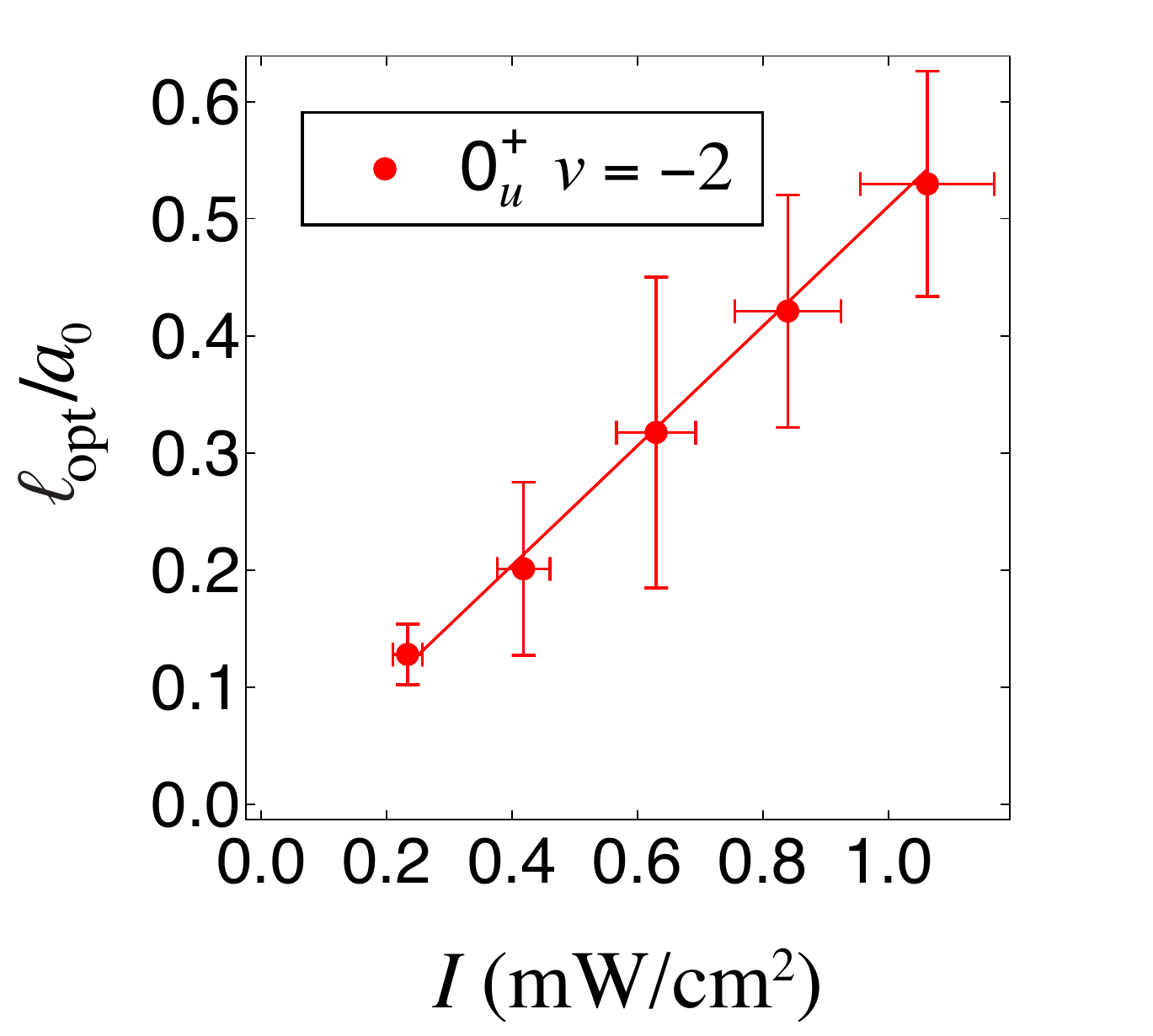}
\end{subfigure}
\begin{subfigure}{0.235\textwidth}

\includegraphics[width=1\linewidth]{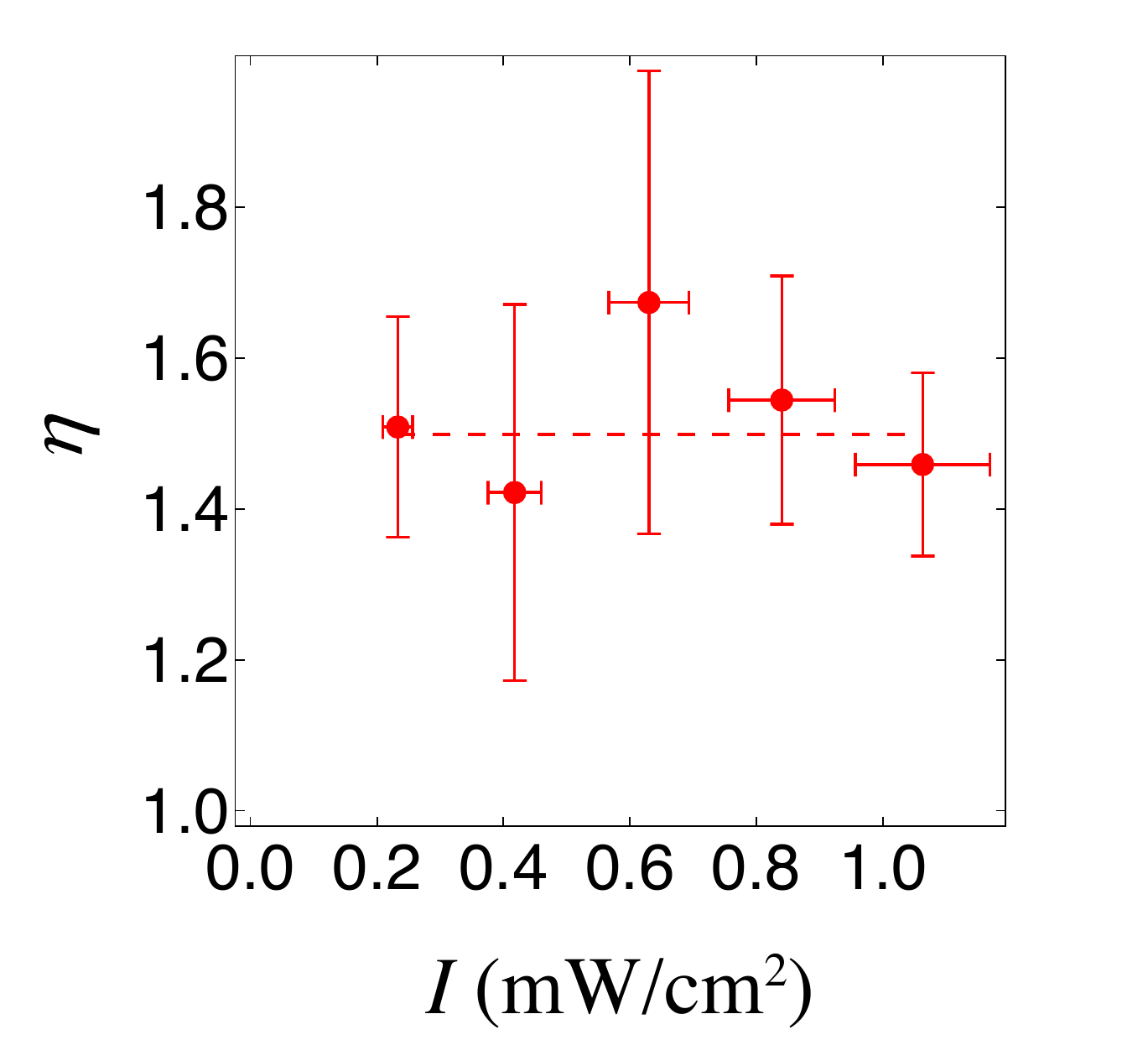}
\end{subfigure}

\begin{subfigure}{0.235\textwidth}

\includegraphics[width=1\linewidth]{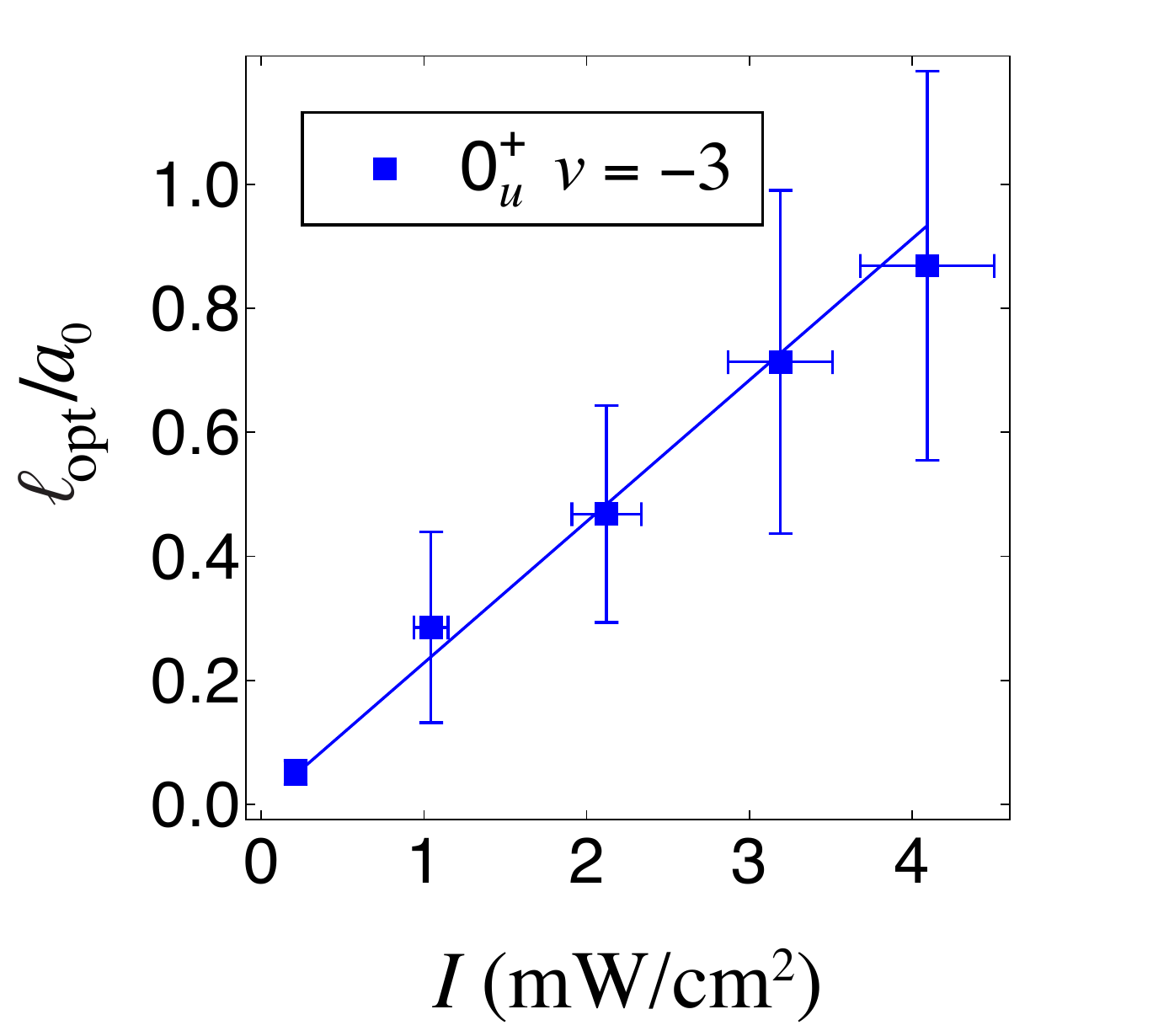}
\end{subfigure}
\begin{subfigure}{0.235\textwidth}

\includegraphics[width=1\linewidth]{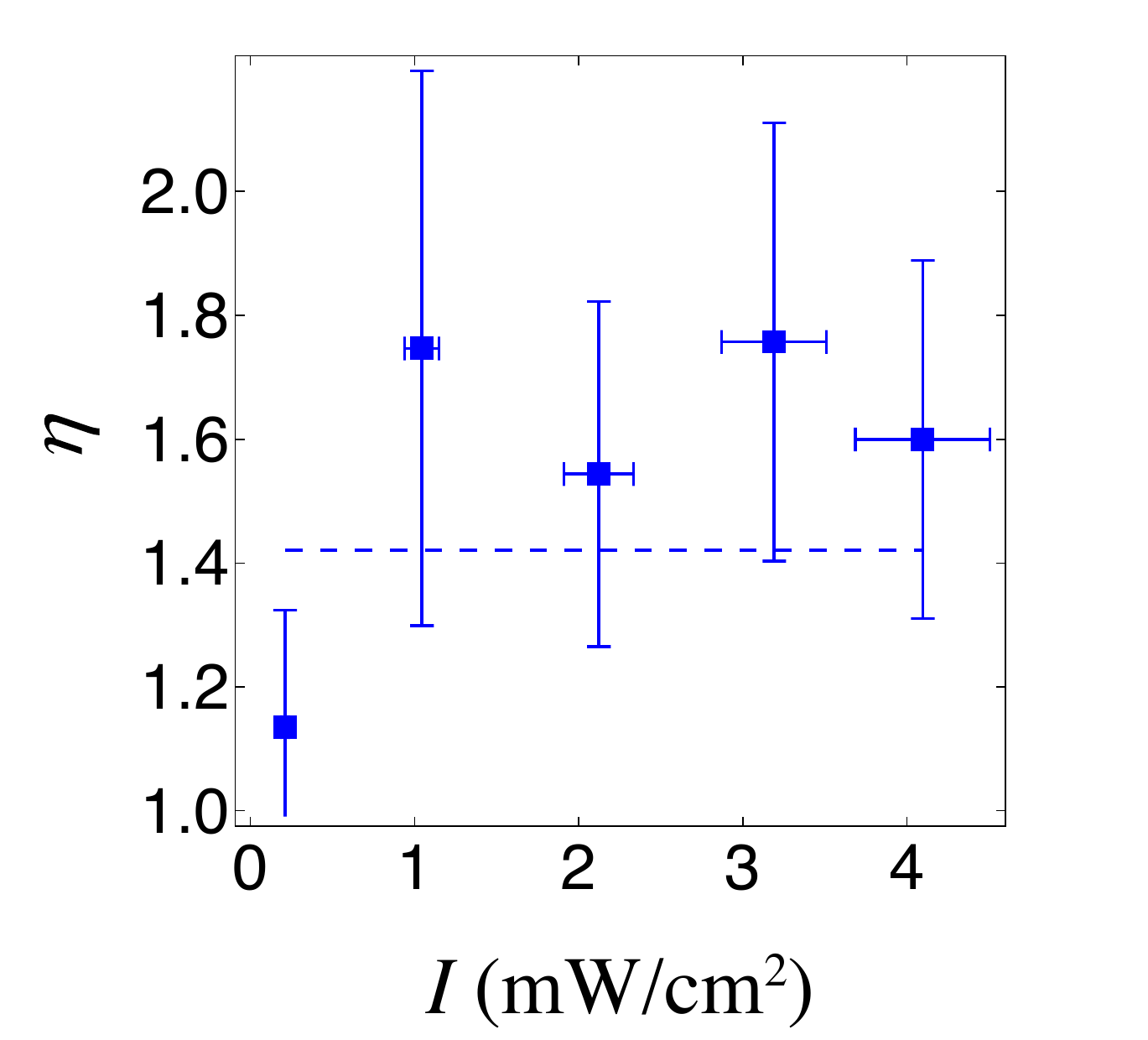}
\end{subfigure}

\begin{subfigure}{0.235\textwidth}
\includegraphics[width=1\linewidth]{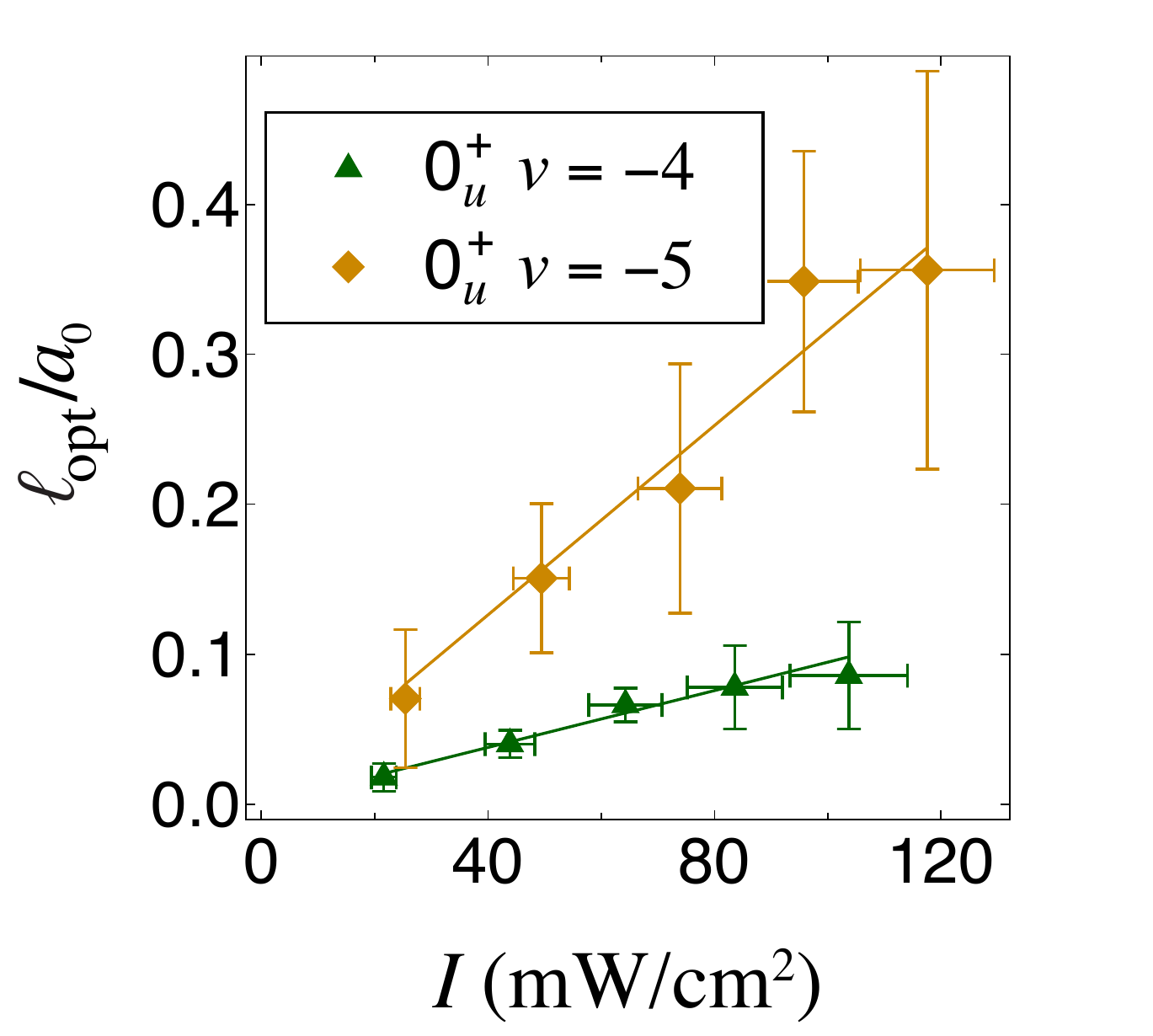}
\end{subfigure}
\begin{subfigure}{0.235\textwidth}

\includegraphics[width=1\linewidth]{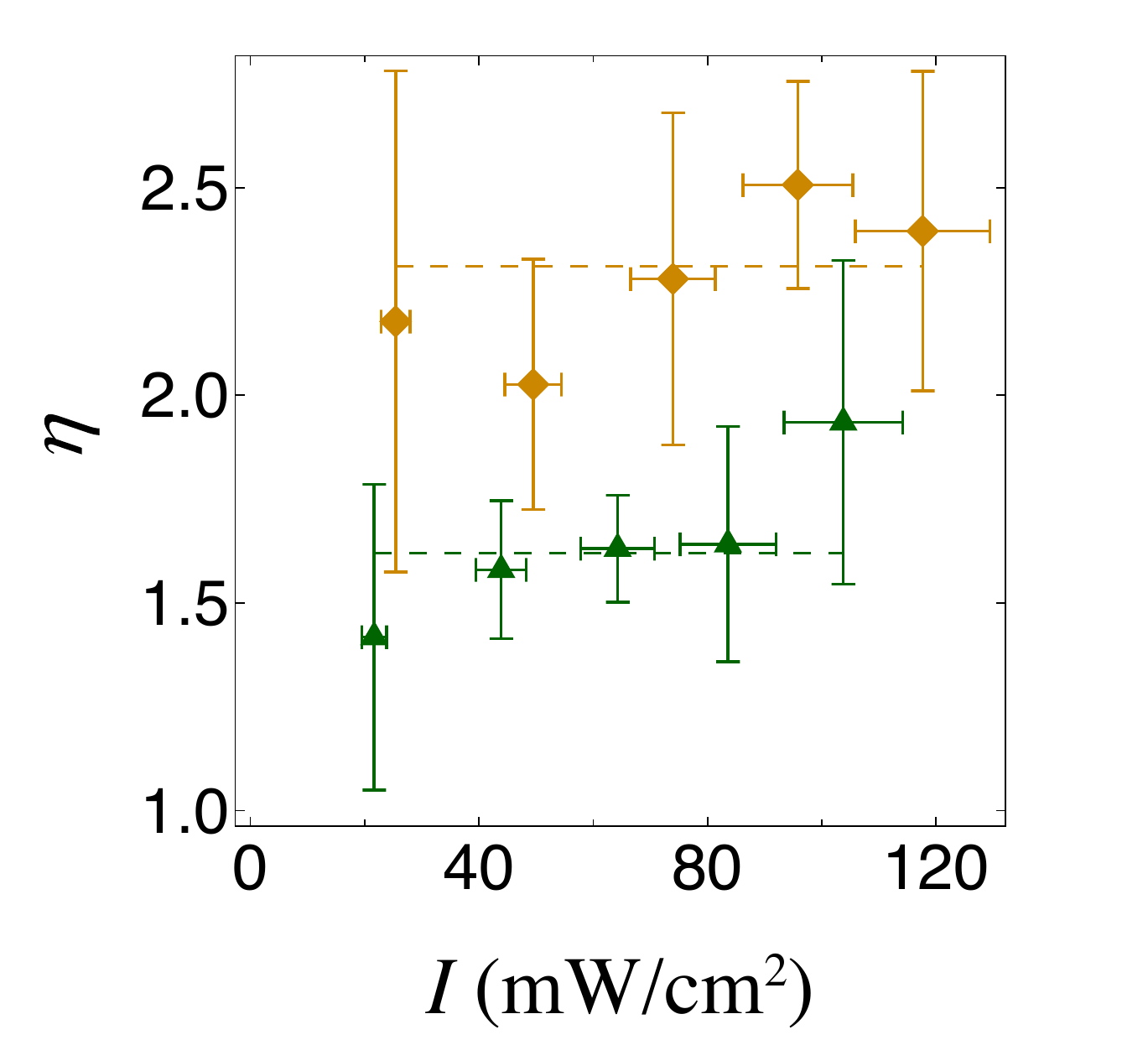}
\end{subfigure}

\caption{\label{fig:OFRresults}(Color online) Optical lengths and $\eta$ broadening factors for the $^{84}$Sr $0_u^+$, $\nu = -2$ to $\nu=-5$ resonances extracted by fitting experimental data to Eq. (\ref{eq:PASNumBEC}). Solid lines indicate linear fits whose slope give the reported values for $\lopt/I$. Dashed lines indicate the weighted average values of $\eta$ for each resonance. Error bars include both systematic and statistical sources as described in Sec.~\ref{sec:loptMeas}.}
\end{figure}

The extracted optical lengths and broadening factors for four $\sr{84}$ $0_u^+$ resonances are shown in Fig.~\ref{fig:OFRresults}. 
Due to its proximity to the atomic transition, we did not attempt to measure the $0_u^+,~\nu=-1$ optical length.
For each resonance, $\ell_\mathrm{opt}/I$ is extracted from the slopes in the left plots and $\eta$ is extracted from the weighted average of points in the right plots. 
We consider systematic error sources such as uncertainties in the dipole trap frequencies, atomic scattering length, measured laser intensity, and measured laser beam size. 
These contributions are added in quadrature with the estimated standard deviation in the fitting parameters to arrive at the total error. 
The largest error source is the statistical uncertainty, followed by systematic uncertainty in measuring the laser intensity.

We calculated theoretical $\ell_\mathrm{opt}/I$ values to compare to the measured values, as reported in Table~\ref{tab:OptLenTesults}.
The theory values are based on numerical calculations of the ground and excited molecular wavefunctions. 
For the ground state scattering potential we used a Lennard-Jones potential of the form
\begin{equation}
V(r) = \frac{C_{12}}{r^{12}} - \frac{C_{6}}{r^{6}} -\frac{C_{8}}{r^{8}} -\frac{C_{10}}{r^{10}}.
\end{equation} 
The long-range $C_6$, $C_8$, and $C_{10}$ coefficients are taken from \cite{Stein2010} and the repulsive $C_{12}$ \cite{Ciurylo2006} term was chosen to reproduce the $\sr{84}$ experimentally determined scattering length ($a_{bg}=123\,a_0$ \cite{MartinezdeEscobar2008}) and potential depth ($\simeq 1081.6$~cm$^{-1}$ \cite{Stein2010}).
We used the Numerov method to numerically integrate the Schr\"odinger equation with this potential and normalized the resulting wavefunction by matching its long-range behavior to the energy-normalized form of Eq.~(\ref{eq:energyNormalizedWF}) using the methods described in Ref.~\cite{Ruzic2013}.
The excited-state molecular wavefunctions are generated by the calculations described in Sec.~\ref{sec:theory}.
The measured and calculated optical lengths match to better than 50~\%. 
The disagreement could be due to systematic errors in the photoassociation atomic loss rate, photoassociation laser intensity and/or theoretical model.

\begin{table}
\centering
\caption[Optical lengths for $^{84}$Sr]{\label{tab:OptLenTesults}Measured $^{84}$Sr optical lengths and broadening factors for four $0_u^+$ resonances. The $\ell_\mathrm{opt}/I$ measurements agree with a numerical estimate to better than 50~\%.}
\begin{ruledtabular}
\begin{tabular}{c d{7} d{5} c c}
$\nu$ & \multicolumn{1}{c}{$E_b/h$ (MHz)}  & \multicolumn{1}{c}{$\eta$} & \multicolumn{2}{c}{$\ell_\mathrm{opt}/I$ ($a_0$(W/cm$^2)^{-1})$} \\[0.5ex]
\cline{4-5} \T
 & & & \multicolumn{1}{c}{Expt.} & \multicolumn{1}{c}{Theory} \\[0.5ex]
\hline \T
$-2$ & -23.050(10) & 1.50(8) & 510(61) & 275 \\[0.5ex]
$-3$ & -228.406(10) & 1.42(12) & 228(42) & 149\\[0.5ex]
$-4$ & -1143.161(10) & 1.62(9) & 0.95(12) & 0.95 \\[0.5ex]
$-5$ & -3692.645(10) & 2.31(15) & 3.16(52) & 3.8 \\[0.5ex]
\end{tabular}
\end{ruledtabular}
\end{table}

Measurements and calculations confirm that the line strength for the $\sr{84}$,~$\nu = -2$ resonance is suppressed with respect to the resonances in $\sr{86}$ and $\sr{88}$ with the same index, for which $\ell_\mathrm{opt}/I \approx 10^4~a_0 (\mathrm{W/cm}^2)^{-1}$ \cite{Borkowski2014,Blatt2011}.
This is because in $\sr{84}$, the Condon point for this resonance ($R_c \simeq 150\, a_0$) is close to the background scattering length $123\, a_0$ and therefore a node in the scattering wavefunction \cite{Takasu2012,Julienne1996}.
The optical lengths of the $\nu=-4$ and $-5$ resonances are also suppressed by nodes in the ground state scattering wavefunction near the classical turning points.
Contrary to the typical trend, the $\nu=-4$ state has a smaller $\ell_\mathrm{opt}/I$ than the $\nu=-5$ state, which is also reproduced by the numerical calculation.

We observe that the $\eta$ broadening factors are independent of laser intensity, which matches the behavior from a similar measurement in $^{174}$Yb \cite{Kim2016}.
The source of $\eta$ is not clear. 
There could be some contribution to the broadening from systematic sources such as laser frequency drift and magnetic field noise.
Another potential broadening source is Doppler broadening due to BEC excitations from the dipole trap modulation discussed in Sec.~\ref{sec:exp}.
Though the dipole traps are modulated with a frequency ($f_\mathrm{mod}=2$~kHz) much faster than the largest harmonic trapping frequency ($f_\mathrm{trap} \approx 140$ Hz), they will still create some motional excitations in the BEC.
To estimate the magnitude of this effect, we calculated the average atomic velocity using the approach of Ref.~\cite{Castin1996} and determined the typical Doppler broadening to be $\leq 1$~kHz. 
The residual broadening is likely due to additional molecular loss processes, similar to those that have been measured in other experiments \cite{Theis2004,Zelevinsky2006,Blatt2011,Yan2013b,Borkowski2014,Kim2016}, but lack a theoretical explanation.

\section{\label{sec:conclusion}Conclusions}

In conclusion, we have measured seven photoassociation resonances relative to the $\term{1}{S}{0} + \term{3}{P}{1}$ dissociation limit for $\sr{86}$ and $\sr{84}$.
In addition to confirming several previously measured lines, we also found several new molecular states and updated the measurement of three lines from previous studies.

Using this improved binding energy spectrum, we have developed a theoretical mass-scaled model that reproduces the experimental results to within 1~MHz.
Our results are consistent with the observation from \cite{Borkowski2014} that the $\sr{84}$ $0_u^+$ spectrum is strongly perturbed by the $\term{1}{S}{0} + \term{1}{D}{2}$ $0_u^+$ $(^1\Sigma_u^+)$ potential. 
This improved theoretical treatment of the strontium bound states will aid future efforts to produce ground state molecules or use these lines as OFRs.

In addition, these results may inform future photoassociation experiments with the fermionic isotope of strontium, $\sr{87}$.
As shown in a similar study of $^{173}$Yb, resonances corresponding to different total angular momentum are likely to be individually resolved \cite{Han2018}.
Therefore, it may be possible to engineer spin-state dependent optical
Feshbach resonances. 
Such a tool may be useful in quantum simulation schemes as a
method of controllably breaking the \textit{SU(N)} symmetry of the system.

The authors thank Mateusz Borkowski, Eite Tiesinga, and Dimitrios Trypogeorgos for useful discussions as well as Peter Elgee for experimental assistance.
This work was partially supported by the U.S.\ Office of Naval Research, and the NSF through the Physics Frontier Center at the Joint Quantum Institute.

\bibliography{SrPAS,SrPAS2}

\end{document}